\documentclass[pra,showpacs,graphics,twocolumn,floatfix,mathbbm,nofootinbib]{revtex4-1}
\usepackage{amsthm}
\usepackage{amsmath}
\usepackage{latexsym}
\usepackage{amsfonts}
\usepackage{amssymb}
\usepackage{color}
\usepackage{bbm,dsfont}
\usepackage{graphicx}
\usepackage[caption=false]{subfig}
\usepackage{mathrsfs}
\usepackage{mathbbol}
\usepackage{hyperref}
\usepackage{enumerate}
\usepackage{MnSymbol}
\usepackage{bbding}
\usepackage{ulem}

\newtheorem{proposition?}{Proposition?}

\theoremstyle{definition}

\newcommand{\ket}[1]{|#1\rangle} 
\newcommand{\bra}[1]{\langle#1|} 

\begin{document}
\title[]{Entanglement Protection via Periodic Environment Resetting \\ in continuous-time Quantum Dynamical Processes }
 
\author{Thomas Bullock}
\affiliation{QTF Centre of Excellence, Turku Centre for Quantum Physics, Department of Physics and Astronomy, University of Turku, FI-20014 Turun yliopisto, Finland}

\author{Francesco Cosco}
\affiliation{QTF Centre of Excellence, Turku Centre for Quantum Physics, Department of Physics and Astronomy, University of Turku, FI-20014 Turun yliopisto, Finland}

\author{Marwan Haddara}
\affiliation{QTF Centre of Excellence, Turku Centre for Quantum Physics, Department of Physics and Astronomy, University of Turku, FI-20014 Turun yliopisto, Finland}

\author{Sina Hamedani Raja}
\affiliation{QTF Centre of Excellence, Turku Centre for Quantum Physics, Department of Physics and Astronomy, University of Turku, FI-20014 Turun yliopisto, Finland}

\author{Oskari Kerppo}
\affiliation{QTF Centre of Excellence, Turku Centre for Quantum Physics, Department of Physics and Astronomy, University of Turku, FI-20014 Turun yliopisto, Finland}

\author{Leevi Lepp\"aj\"arvi}
\affiliation{QTF Centre of Excellence, Turku Centre for Quantum Physics, Department of Physics and Astronomy, University of Turku, FI-20014 Turun yliopisto, Finland}

\author{Olli Siltanen}
\affiliation{QTF Centre of Excellence, Turku Centre for Quantum Physics, Department of Physics and Astronomy, University of Turku, FI-20014 Turun yliopisto, Finland}

\author{N. Walter Talarico}
\affiliation{QTF Centre of Excellence, Turku Centre for Quantum Physics, Department of Physics and Astronomy, University of Turku, FI-20014 Turun yliopisto, Finland}

\author{Antonella De Pasquale}
\affiliation{Department of Physics and Astronomy, University of Florence, Via G. Sansone 1, 50019, Sesto Fiorentino (FI), Italy}
\affiliation{INFN Sezione di Firenze, via G. Sansone 1, I-50019 Sesto Fiorentino (FI), Italy}
\affiliation{NEST, Scuola Normale Superiore and Istituto Nanoscienze-CNR, I-56127 Pisa, Italy}

\author{Vittorio Giovannetti}
\affiliation{NEST, Scuola Normale Superiore and Istituto Nanoscienze-CNR, I-56127 Pisa, Italy}

\author{Sabrina Maniscalco}
\affiliation{QTF Centre of Excellence, Turku Centre for Quantum Physics, Department of Physics and Astronomy, University of Turku, FI-20014 Turun yliopisto, Finland}
\affiliation{QTF Centre of Excellence, Department of Applied Physics, Aalto University, FI-00076 Aalto, Finland}

\begin{abstract}
The temporal evolution of entanglement between a noisy  system and an
ancillary system is analyzed in the context of continuous-time open quantum system dynamics. Focusing on a couple
of analytically solvable models for qubit systems, 
we study how Markovian and non-Markovian characteristics influence the problem, 
discussing in particular their associated  entanglement-breaking regimes. 
These performances are compared with those one could achieve when the environment of the system is 
forced to return to its input configuration via periodic instantaneous resetting procedures. 
\end{abstract}

\maketitle

\section{Introduction}

The preservation of entanglement  is a fundamental requirement for the development of realistic applications in quantum communication~\cite{COMM1}, quantum computation~\cite{NIELSEN}, and quantum cryptography~\cite{CRYPTO}. According to quantum mechanics, two entangled systems exhibit 
extraordinary, but fragile, correlations that are beyond any classical description~\cite{ENT}. A key objective in the development of
reliable quantum technologies  is to identify strategies which would prevent the deterioration of such exotic correlations. A plethora of different strategies have been devised to tackle this delicate issue, spanning from distillation protocols~\cite{DIST},
 pre- and postprocessing operations~\cite{ERR1,ERR2,ERR3,ERR4}, decoherence-free subspaces~\cite{DEC,DEC2},
to dynamical decoupling and control techniques~\cite{DYNA1,DYNA2,DYNA3,FacchiLidar,maniscalco-2008,cuevas-2017,
 CONTROL1}.
All these approaches, however, are uneffective if  the noise level affecting  the system surpasses a certain minimal threshold
 that leads to entanglement-breaking (EB) dynamics~\cite{horodecki-2003}. 
 A quantum  process is said to be EB if  any initial amount of entanglement established between the system, evolving under the
 action of the noise, and an arbitrary external ancilla is destroyed. 
 Such transformations behave essentially as classical measure-and-reprepear operations~\cite{M&R} and have been the subject of extensive investigation within the quantum information community; see, e.g., Refs.\cite{horodecki-2003,chruscinski-2006}.
 Determining when the EB threshold is approached during a given dynamical evolution
 is clearly an important facet for the construction of
 procedures which are more effective in the protection of quantum coherence;
see, e.g., Ref.~\cite{Gatto18}. 
The present work focuses on this task by studying the continuous-time evolution of a qubit whose dynamics are described by generalized master equations that admit explicit integration and allow one to probe both Markovian and non-Markovian
regimes. 
While justified in certain  contexts, the Markov approximation fails when the system-environment interaction leads to long-lasting and non-negligible correlations \cite{breuer-2002}. 
Indeed, in general, the dynamics of an open system are non-Markovian~\cite{NMDOQS, ReviewRHP},  i.e., they are affected by memory effects which may lead to a reappearance of entanglement after its disappearance \cite{bellomo-2007,mazzola2009}. This nonmonotonic behavior of the entanglement between a system and an ancilla was indeed proposed in Ref. \cite {NMRHP} as a non-Markovianity witness. Generally speaking, it can also happen that entanglement revives a long time after its sudden death or that, due to correlations with the environment, the phenomenon of entanglement trapping occurs \cite{enttrap}, resulting in a highly  nontrivial temporal dependence of the entanglement evolution. 

In the first part of the paper we shall review the above effects,  linking them to the  EB analysis of the system dynamics.  
We then proceed by introducing a method which, with minimal control on the system environment, allows one to substantially modify the EB dynamical response. The scheme we propose takes inspiration from 
 the notion of amendable channels introduced in Ref.~\cite{anto} and subsequently developed in Refs.~\cite{depasquale-2013,CUEVAS1,porzio,cuevas-2017}. Formally speaking, an amendable channel  is an EB process resulting from the temporal concatenation  of a collection of subprocesses,  such that there exists  a (typically unitary) filtering transformation acting on the system of interest in a bang-bang control fashion, which applied between the subchannels enables one to create a new effective evolution  which is no longer EB. In our case we adapt this idea to the continuous-time evolution of a qubit,  by assuming that we indirectly perturb its dynamics via periodic, instantaneous  
resettings of its environment. 
It is worth stressing that, at odds with the approaches of Refs.~\cite{cuevas-2017,anto,depasquale-2013,CUEVAS1,porzio},
 our scheme  assumes
partial control on the environment, which, in general, may not be granted.
Still there are several  reasons to study this procedure.  First, there are configurations where the resetting assumption is an available option. For instance, in the amplitude-damping scheme 
describing the interaction of a two-level system $A$ with a
 bosonic reservoir at  zero temperature (a model we analyze in Sec.~\ref{TLADP}), the resetting merely accounts for periodically removing all the excitations
 that leaked out from $A$ (e.g., by means of ancillary systems that act as effective photonic sinks),
 or by preventing them from being reabsorbed from $A$ (e.g., by instantaneously detuning the latter). 
Second, the environment-resetting assumption is interesting because, despite the fact that it explicitly 
contrasts the back-flow of information from the bath to $A$, thereby apparently  increasing the overall noise level of the dynamics, in certain regimes it  allows us to improve the  entanglement survival time of the model. 
 Finally, from a mathematical point of view the resetting assumption  results in a huge 
simplification of  the problem ,as without it, it would be impossible to 
write the perturbed evolution of the system in a compact, treatable form, at least for non-Markovian processes.

The paper is structured as follows. 
In Sec.~\ref{SECii}  we  review the formal definitions
of non-Markovian quantum-dynamical processes and of EB quantum maps. In Sec.~\ref{SECiii} we present a couple of dynamical 
processes for qubit systems and discuss their EB properties.
The environment-resetting procedure is presented in Sec.~\ref{SEC:4}. Finally, we discuss our results and illustrate their possible connection with the quantum Zeno and inverse quantum Zeno effects in Sec.~\ref{CONCLUSIONS}. 
Technical material is presented in the appendices. 

\section{Definitions}\label{SECii}
In this section we review some basic facts about open quantum system dynamics and their characterization in terms of Markovian and non-Markovian models,  and introduce the formal definition of EB channels.
\subsection{Continuous-time open quantum processes} \label{SECiiA}
 In the continuous-time approach the dynamics of an open quantum system $A$ are described by a $t$-parametrized family $\{\Phi_{0 \rightarrow t}\}_{t\geq 0}$ of completely positive (CP), trace-preserving maps (quantum channels) that link a generic input state $\varrho^A(0)$ of the system  to its  temporal evolved counterpart $\varrho^A(t)$  via the mapping 
 \begin{eqnarray} \varrho^A(t) = \Phi_{0 \rightarrow t}(\varrho^A(0))\;. \label{DEFMAP} \end{eqnarray} 
In this setting,  time-homogenous Markovian dynamics can be  associated to the semigroup property 
\begin{eqnarray} 
\Phi_{0\rightarrow t}=\Phi_{0\rightarrow t-s}  \circ \Phi_{0\rightarrow s}\;, \qquad \mbox{$\forall t$ and $\forall s \leq t$,} \label{DEFPRO0}
\end{eqnarray}
with ``$\circ$"  representing the composition of maps, i.e., $ \Phi_{0\rightarrow t-s}  \circ \Phi_{0\rightarrow s}(\varrho)= \Phi_{0\rightarrow t-s}( \Phi_{0\rightarrow s}(\varrho))$. 
Equivalently, we can associate these dynamics to the Gorini-Kossakowski-Sudarshan-Lindblad (GKSL) form of the master equation~\cite{Lindblad, GKS}, which describes the evolution of 
the system's density matrix in terms of time-independent Hamiltonian  $H$ and dissipator ${\cal D}$, 
\begin{eqnarray} \label{ME} 
\frac{\partial}{\partial t} \varrho^A(t) &=& - i [ H, \varrho^A(t) ] + {\cal D}(\varrho^A(t))\;, \\ 
  {\cal D}(\cdot) &=& \sum_j \left(L_j \cdot L_j^\dag - \frac{1}{2} \{ L_j^\dag L_j,  \cdot \}\right) \;, 
\end{eqnarray}
with $[\cdot,\cdot]$ and $\{ \cdot, \cdot\}$ being the commutator and anticommutator, respectively, and $L_j$ the Lindblad operators. 
 
 Starting from a microscopic model of the system, environment, and interaction, the enforcement of condition \eqref{DEFPRO0} 
  on the system's dynamics 
 requires a number of assumptions, such as system-reservoir weak coupling~\cite{breuer-2002}. In certain physical contexts, however, such approximations are unjustified, and one needs to go beyond perturbation theory. A straightforward extension of the  semigroup property \eqref{DEFPRO0}  is the notion of divisibility.   A dynamical map is CP-divisible, or simply divisible, iff the propagator $\Lambda_{s\rightarrow t}$ defined through the expression
\begin{eqnarray} 
\Phi_{0\rightarrow t}= \Lambda_{s\rightarrow t}\circ \Phi_{0\rightarrow s}\;, \label{DEFPRO}
\end{eqnarray}
is CP, $\forall t$ and $\forall s \leq  t$. 
 This amounts to saying that the evolution of $A$ can be described at all times as a concatenation of quantum channels, a condition which allows us to still write 
 a linear differential equation for $\varrho^A(t)$ as in \eqref{ME} with explicitly time-dependent  operators $H(t)$ and $L_j(t)$. 
Of course, systems obeying \eqref{DEFPRO0} can be seen as special instances of divisible models with CP propagators $\Lambda_{s\rightarrow t}$ that also fulfill the constraint 
\begin{eqnarray}\label{TRANS} 
\Lambda_{s\rightarrow t} = \Lambda_{0\rightarrow t-s }\;, \qquad \mbox{$\forall t$ and $\forall s \leq t$,}
\end{eqnarray} 
which explains why we dubbed them as  ``time-homogenous Markovian"  instead of simply Markovian processes.  Any process whose dynamics are not divisible are considered non-Markovian.

A non-Markovianity measure quantifying the deviation from divisibility
 has been proposed in~\cite{NMRHP}
whose physical interpretation has only very recently been  fully unveiled \cite{BognaToni}. 
Specifically Ref.~\cite{NMRHP} introduces a witness of nondivisibility which exploits the temporal evolution  of the entanglement  between the open system state and an external ancilla: a nonmonotonic decay of such entanglement indicates nondivisibility, and therefore non-Markovianity, of the dynamical map. 
A different perspective on the definition of non-Markovianity is to interpret memory effects in terms of information back-flow. This path was first undertaken by Breuer, Laine, and Piilo by quantifying the information content of an open quantum system in terms of distinguishability between pairs of states \cite{NMBLP}. Several other information-theoretic measures of non-Markovianity have been proposed in the last decade \cite{ReviewRHP,NMDOQS}. The key property exploited in these definitions is that the time evolution of a given quantifier of information suitable to describe memory effects, e.g., distinguishability between quantum states, is contractive under CP maps. Hence a temporary increase of distinguishability, which is physically interpreted as a partial increase in the information content of the open system due to memory effects, always implies that divisibility of the dynamical map is violated.

\subsection{{Entanglement breaking channels}}

A quantum channel $\Phi$ acting on a system $A$ is said to be EB~\cite{horodecki-2003} if,
irrespective of the choice of joint state $\varrho^{AB}$ of $A$ and of an arbitrary ancilla $B$,  the associated output $(\Phi \otimes {\rm Id})(\varrho^{AB})$ is separable,  ${\rm Id}$ indicating the identity channel on $B$. For finite dimensional systems the Choi-Jamio{\l}kowski isomorphism \cite{choij,JAM} allows us to restrict the analysis to the case where 
$B$ is isomorphic to $A$ and  $\varrho^{AB}$ is the maximally entangled state 
$|\Omega\rangle_{AB}=\frac{1}{\sqrt{d}} \sum_{k=1}^d \ket{k}_A\!\otimes\ket{k}_B$,
$\{|k\rangle\}_{k=1,\cdots, d}$ being an orthonormal basis on the associated Hilbert space. Such an output density matrix 
$\varrho^{AB}_{\Phi} = (\Phi \otimes {\rm Id})(|\Omega\rangle_{AB}\langle \Omega|)$ is called the
Choi-Jamio{\l}kowski (CJ) state of $\Phi$, 
and  its separability is equivalent to the EB property of the map. 
In what follows we shall focus on the case where $A$ is a qubit system. Accordingly we identify $\ket{\Omega}_{AB}$ with the Bell state 
 $\tfrac{1}{\sqrt 2}(\ket{00} + \ket{11})$,  and use the concurrence~\cite{wootters-1998} 
 of the CJ state 
  as a necessary and sufficient instrument to determine whether or not the associated map is EB. 
  We remind the reader that given a two-qubit state $\varrho^{AB}$, its concurrence  ${C}[\varrho^{AB}]\in[0,1]$
 is a proper entanglement measure which assumes nonzero values if and only if  $\varrho^{AB}$ is entangled. It  can be computed as \begin{equation}
	C[\varrho^{AB}]= \max\{0, \sqrt{\varepsilon_1}-\sqrt{\varepsilon_2}-\sqrt{\varepsilon_3}-\sqrt{\varepsilon_4}\},
\end{equation}
where $\{\varepsilon_i\}_{i=1}^4$ is the set of eigenvalues (in descending order) of the operator
	$\chi = \varrho^{AB}(\sigma_2^A \otimes \sigma_2^B) \varrho^{AB*}(\sigma_2^A \otimes \sigma_2^B)$,
with $\varrho^{AB*}$ the complex conjugate of  $\varrho^{AB}$ and $\sigma^X_2$ being the second Pauli matrix acting
on the system $X=A,B$. 

\section{EB properties of dynamical processes acting on a qubit} \label{SECiii} 

{In this section we present a couple of examples of dynamical processes $\{ \Phi_{0\rightarrow t}\}_{t\geq 0}$ for a qubit system
which are exactly solvable and which, depending on the model parameters, allow one to describe both Markovian and non-Markovian 
evolutions.
In particular we are interested in studying their EB properties as a function of the temporal index $t$. According to the 
previous section this can be done by looking at the zeros of the concurrence $C(t)$ of the  CJ state 
\begin{eqnarray} \varrho^{AB}_{\Phi_{0\rightarrow t}}
=({\Phi_{0\rightarrow t}} \otimes \mathrm{Id})(|\Omega\rangle_{AB}\langle \Omega|) \label{CJSTATE1}  
\end{eqnarray} 
of the map $\Phi_{0\rightarrow t}$, i.e., by
solving the equation 
\begin{eqnarray}\label{NEWCON} 
 C(t) ={C}[ \varrho^{AB}_{\Phi_{0\rightarrow t}}]=0\;.
 \end{eqnarray}  
 For divisible  processes, due to the CP property of the  propagator $\Lambda_{s\rightarrow t}$ of Eq.~\eqref{DEFPRO},  the function 
 $C(t)$ is explicitly nonincreasing. Therefore, 
after the concurrence  reaches zero, it remains that value 
for  all subsequent instants ~\cite{Gatto18}. By contrast,
in the general non-Markovian setting this is not necessarily true as the associated function $C(t)$ can be explicitly nondecreasing 
due to information back-flow. However, notice that, as previously mentioned, since
the nonmonotonic behavior of entanglement measures is only a witness of non-Markovianity~\cite{NMRHP}, there exist 
non-Markov processes which still admit nonincreasing $C(t)$, e.g., when the propagator  $\Lambda_{s\rightarrow t}$ of the family 
  is just  positive but not CP~\cite{PDIVISIBILITY}. }

\subsection{Time-local amplitude-damping channels}\label{TLADP} 

As a first case study we consider an amplitude-damping channel for a two-level atom (qubit) $A$,
whose density matrix evolves according to the 
 time-local differential master equation 
\begin{eqnarray}
	\dfrac{d\varrho^A(t)}{dt} = \gamma(t)\left( \sigma_- \varrho^A(t) \sigma_+ - \dfrac{1}{2} \{\sigma_+	\sigma_-, \varrho^A(t) \} \right), \label{eq:ampldamp}
\end{eqnarray}
where $\sigma_{\pm}= \frac{1}{2}(\sigma_1 \pm i \sigma_2)$ are the raising and lowering operators of the system. 
In this expression the function $\gamma(t)$ is an effective (time-dependent) rate, which, as will become clear in a moment,  need not be positive semidefinite at all times. 
Equation~\eqref{eq:ampldamp} admits an analytical integration whose solution, expressed in the  eigenbasis $\{ |0\rangle_A, |1\rangle_A\}$ of the $\sigma_3$ Pauli operator,
results in 
\begin{equation}\label{eq:ad-1q}
	\varrho^A(t) = \begin{pmatrix}
	\varrho^A_{11}(0) P(t) & \varrho^A_{10}(0) \sqrt{P(t)} \\
	\varrho^A_{01}(0) \sqrt{P(t)} & \varrho^A_{00}(0) + \varrho^A_{11}(0)(1-P(t))
	\end{pmatrix},
\end{equation}
with 
\begin{eqnarray} 
P(t) := e^{ -\Gamma(t)}, \label{defP}   \qquad \Gamma(t) := \int_0^t dt' \gamma(t')\;, 
\end{eqnarray} 
representing the population of the level $|1\rangle_A$. The above expressions clarify the condition that the rate $\gamma(t)$ has to fulfill in order to 
interpret Eq.~\eqref{eq:ad-1q} as an instance of Eq.~\eqref{DEFMAP} for a proper choice of the quantum channel~$\Phi_{0\rightarrow t}$:
indeed, exploiting the fact that a necessary and sufficient CP condition for \eqref{eq:ad-1q} is to have the function $P(t)$ be positive and no larger than 1, 
it follows that Eq.~\eqref{eq:ampldamp} is a legitimate dynamical equation for $A$ if and only if  the function $\gamma(t)$ respects the constraint 
\begin{eqnarray}
\Gamma(t) \geq 0 \;, \qquad \forall t\geq 0\;. \label{COND111} 
\end{eqnarray} 
Equation~\eqref{COND111} is clearly fulfilled if we enforce the positivity condition directly on $\gamma(t)$. 
Under this restriction Eq.~\eqref{eq:ampldamp} is explicitly in the generalized  GKSL form, characterized by a single
time-dependent Linbdlad operator $L(t)= \sqrt{\gamma(t)} \sigma_-$, and the resulting  process is divisible. 
Furthermore, if we take the rate  to be positive and constant $\gamma(t)=\lambda$, then the maps $\Phi_{0\rightarrow t}$  become time-homogeneous, yielding 
a  population $P(t)$ which is exponentially decreasing:
\begin{eqnarray} P(t) = e^{- \lambda t} \;. \label{TIMEHO} \end{eqnarray} 
Finally, if $\gamma(t)$  assumes negative values   [while still fulfilling~\eqref{COND111}] the resulting process is non-Markovian as we explicitly show next.

Indeed, by direct evaluation one can verify that the  CJ state~\eqref{CJSTATE1}  of the model has the following 
 $X$-shaped form:
 \begin{equation} \label{CJEX1} 
	\varrho^{AB}_{\Phi_{0\rightarrow t}}=
	\dfrac{1}{2}\begin{pmatrix}
	P(t)& 0 & 0 & \sqrt{P(t)} \\
	0 & 0 & 0 & 0 \\
	0 & 0 & 1-P(t) & 0 \\
	\sqrt{P(t)} & 0 & 0 & 1
	\end{pmatrix},
\end{equation}
 whose concurrence can be explicitly computed~\cite{ERR4} resulting in the expression 
\begin{equation}
	C(t) = \sqrt{P(t)} = e^{ -\Gamma(t)/2} \label{eq:sabb}
\end{equation}
(see Appendix  \ref{CJSTATE} for details).
Taking the derivative and invoking Eq.~\eqref{defP}, 
we now have
\begin{equation}
	\frac{d }{dt} C(t) = - \frac{\sqrt{P(t)}}{2} \gamma(t)  , \label{eq:sab}
\end{equation}
which shows that a negative value of $\gamma(t)$   implies an increasing behavior  for $C(t)$ and, as anticipated, 
a non-Markovian character of the system's dynamics via a direct application of the sufficient condition of Ref.~\cite{NMRHP} . 

\begin{figure}[t!]
\centering
\includegraphics[width=\linewidth]{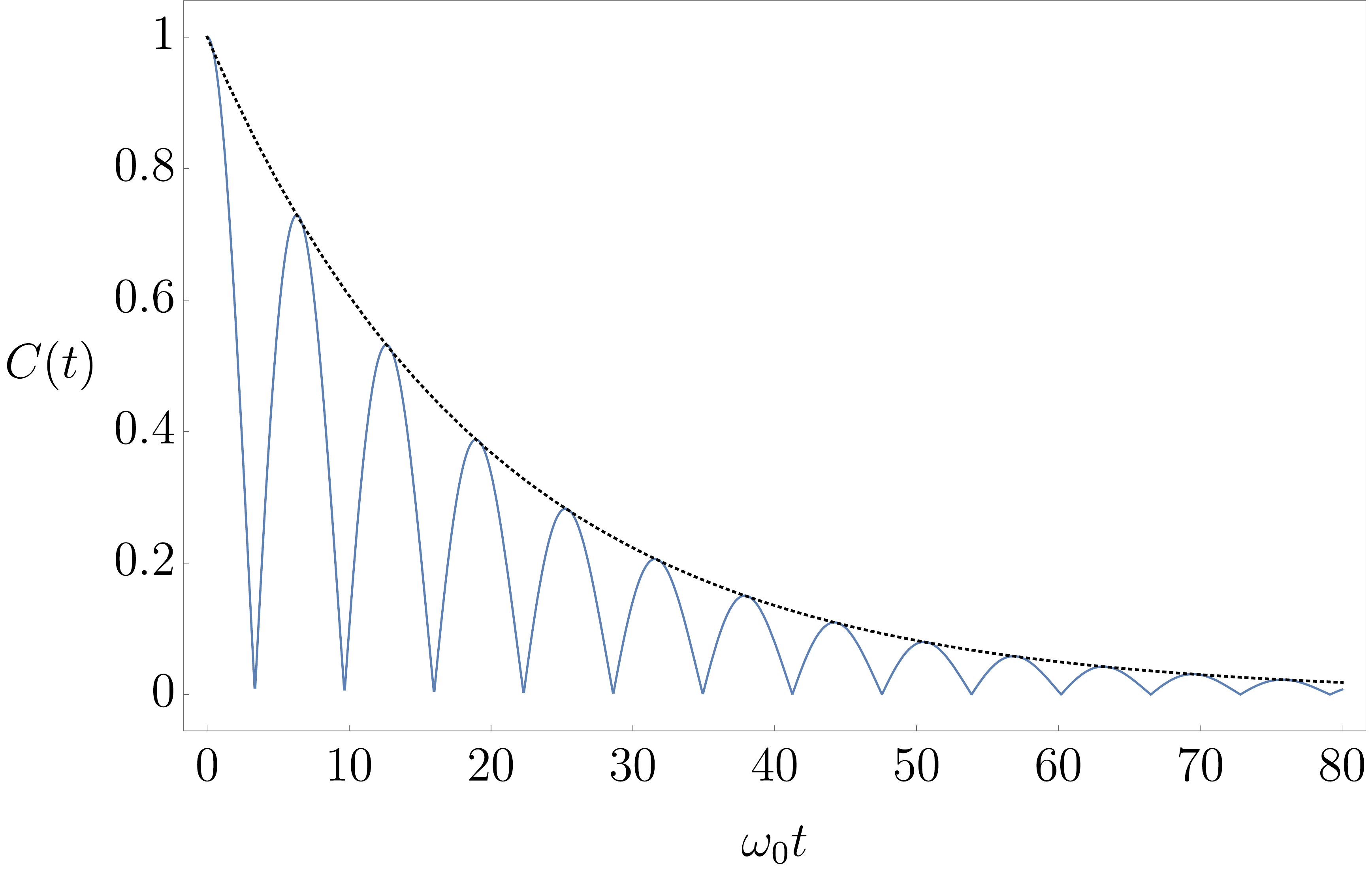}
\caption{\label{fig0}(Color online) Time evolution of the CJ concurrence \eqref{eq:sabb} of the amplitude-damping channel~\eqref{eq:ad-1q} characterized by probability $P(t)$ of Eq.~\eqref{eq:poft}, 
 in the non-Markovian regime with $\alpha = 5 \omega_0$ and $\ell=0.1\omega_0$ (blue solid line). The black dotted line on top corresponds to the  exponential envelope of the curve, i.e., the function $e^{-\ell t/2}$ which incidentally corresponds to the CJ concurrence of the time-homogenous Markovian process of Eq.~\eqref{TIMEHO} with constant rate $\lambda=\ell$.}
\end{figure}

Equation~\eqref{eq:sab}  can also be used to directly link the EB properties of the process to the probability $P(t)$:
in particular we notice that the system becomes EB for those times $t$ where $P(t)$ reaches zero, or equivalently where $\Gamma(t)$
explodes: 
\begin{eqnarray}
\Phi_{0\rightarrow t} \in {\rm EB} \Longleftrightarrow P(t) = 0 \;. 
\end{eqnarray} 
For the case of the time-homogenous Markovian evolution this immediately tells us that the system reaches the EB regime only in the asymptotic
limit $t \rightarrow \infty$.
A less trivial example can be found when studying the interaction of a two-level system $A$ with a
 bosonic reservoir at  zero temperature characterized by 
 a Lorentzian spectral density,
	$J(\omega) := \dfrac{1}{2 \pi} \dfrac{\alpha \ell^2}{(\omega_0 - \omega)^2 + \ell^2}$
with $\alpha\geq 0$ the effective coupling constant, $\ell$ the width of the Lorentzian spectrum, and the frequency $\omega_0$ gauging the energy gap of $A$; see, e.g., Refs.~\cite{breuer-2002,NMBLP}.
Under this condition one can show that the probability $P(t)$ gets expressed as 
\begin{equation} 
	P(t) = e^{-\ell t} \left[ \cos\left(\dfrac{\Delta \; t}{2}\right) + \dfrac{\ell}{\Delta} \sin\left(\dfrac{\Delta \; t}{2} \right) \right]^2\;,  \label{eq:poft}
\end{equation}
with 
\begin{eqnarray} 
\Delta:=\sqrt{\ell (2 \alpha - \ell)}\;.\end{eqnarray} 
A close inspection of these equations reveals that when $\alpha/\ell \leq 1/2$ the excited state probability $P(t)$ decays monotonically to zero, reaching such a value only asymptotically. 
As shown in Ref. \cite{NMBLP}, in this case the information, as measured, e.g., by state distinguishability, flows from system to environment: accordingly, 
in agreement with our previous observation, the dynamics are divisible, and the system becomes EB only at infinite time. In the opposite parameter regime, i.e., when $\alpha/\ell > 1/2$, $P(t)$ has an oscillatory behavior vanishing at times 
 \begin{eqnarray} 
 \tau_k:=\frac {2}{\Delta}(k \pi - \arctan (\Delta/\ell)),
 \label{ZEROS} \end{eqnarray}
 with $k\geq 1$ an integer. 
Memory effects in this case kick in and appear as information back-flow, divisibility is lost, and the dynamics are non-Markovian \cite {NMBLP}. 
Accordingly $C(t)$ acquires an oscillating behavior, periodically reaching zero at the special times~\eqref{ZEROS} where
the process becomes instantaneously EB; see Fig. \ref{fig0}.

\subsection{Time-local Pauli channels}

As a second example of a continuous-time quantum process,  we consider  the case of a qubit evolving under the action of the following time-local Pauli channel, described by the master equation 
\begin{equation} 
\dfrac{d\varrho^{A}(t)}{dt}  = \frac 12 \sum_{k=1}^3 \gamma_k (t) (\sigma_k \varrho^{A}(t) \sigma_k - \varrho^{A}(t) ), \label{eq:Paulit}
\end{equation}
where $\gamma_k(t)$ are time-dependent decay rates
fulfilling the inequalities 
\begin{eqnarray}\label{CPCOND} \Gamma_k(t): = \int_0 ^t \gamma_k (t') dt' \geq 0\;, \qquad \forall k=1,2,3. \end{eqnarray}  
Analogously to Eq.~\eqref{COND111} in  the previous section, the above is a necessary and sufficient condition to guarantee the complete positivity  of the associated dynamical maps $\Phi_{0\rightarrow t}$
of the process, which 
by direct integration can be  expressed as a sum of unitary transformations applied to $A$. Specifically defining 
 $\beta_i(t):=\exp[-\Gamma_j(t)-\Gamma_k(t)]$ for $i\neq j\neq k$ and introducing 
 the functions \begin{align}
	p_1(t) &:= \frac 14 \left[ 1 -\beta_3(t) -\beta_2(t) +\beta_1(t) \right], \\
	p_2(t) &:= \frac 14 \left[ 1 -\beta_3(t) +\beta_2(t) -\beta_1(t) \right], \\
	p_3(t) &:= \frac 14 \left[ 1 +\beta_3(t) -\beta_2(t) -\beta_1(t) \right],
\end{align}
and 	
\begin{eqnarray}
p_0(t) &:=& 1- \sum_{k=1}^3 p_k(t) \nonumber \\
&=&  \frac 14 \left[ 1 +\beta_3(t) +\beta_2(t) +\beta_1(t) \right] ,
\end{eqnarray} 
 we can write 
 \begin{equation}\label{eq:Pauli}
	\Phi_{0\rightarrow t} (\varrho^{A}(0) ) = \sum_{k = 0}^3 p_k (t) \sigma_k \varrho^{A}(0) \sigma_k, 
\end{equation}
where $\sigma_0 = \openone$ is the identity matrix. 
In particular, assuming  the $\gamma_k(t)$ to be equal to a given rate $\gamma(t)$, one has 
\begin{eqnarray}
p_1(t)&=& p_2(t)= p_3(t)=(1- e^{-2\Gamma(t)})/4\;, \\
p_0(t)&=& (1+3 e^{-2\Gamma(t)})/4\;,\end{eqnarray}   and the above equation reduces to 
 \begin{eqnarray}\label{eq:Pauli-uni}
	\Phi_{0\rightarrow t} (\varrho^{A}(0) ) &=& p_0(t)  \varrho^{A}(0)  + \frac{1-p_0(t)}{3}  \sum_{k = 1}^3  \sigma_k \varrho^{A}(0) \sigma_k, \nonumber  \\
	&=& \eta(t)   \varrho^{A}(0)  + \frac{1-\eta(t)}{2} \openone, 
\end{eqnarray}
which describes a qubit depolarizing channel~\cite{NIELSEN,KING}  with noisy parameter 
\begin{eqnarray}   
\eta(t):=\frac{4 p_0(t) -1}{3}
=e^{- 2 \Gamma(t)}\label{NOISY}, 
\end{eqnarray} 
[in deriving this expression we use the identity  $\sum_{k = 0}^3  \sigma_k \varrho^{A}(0) \sigma_k = 2\openone$].

As in the amplitude-damping model,
Eq.~\eqref{eq:Paulit}
allows us to describe different  regimes. In particular
if the $\gamma_k(t)$ are taken to be positive semidefinite, then the associated dynamics are provably divisible, with Eq.~\eqref{eq:Paulit} being explicitly in the GKSL
form with three time-dependent generators $L_k(t) = \sqrt{\gamma_k(t)/2}\; \sigma_k$ (the time-homogeneous limit being reached when further imposing  the rates to be constant). 
Non-Markovian
behaviors can instead be obtained by allowing the rates $\gamma_k(t)$ to assume negative values 
while still respecting the constraint \eqref{CPCOND}. 

\begin{figure}[t!]
\begin{center}
  \includegraphics[width=\linewidth]{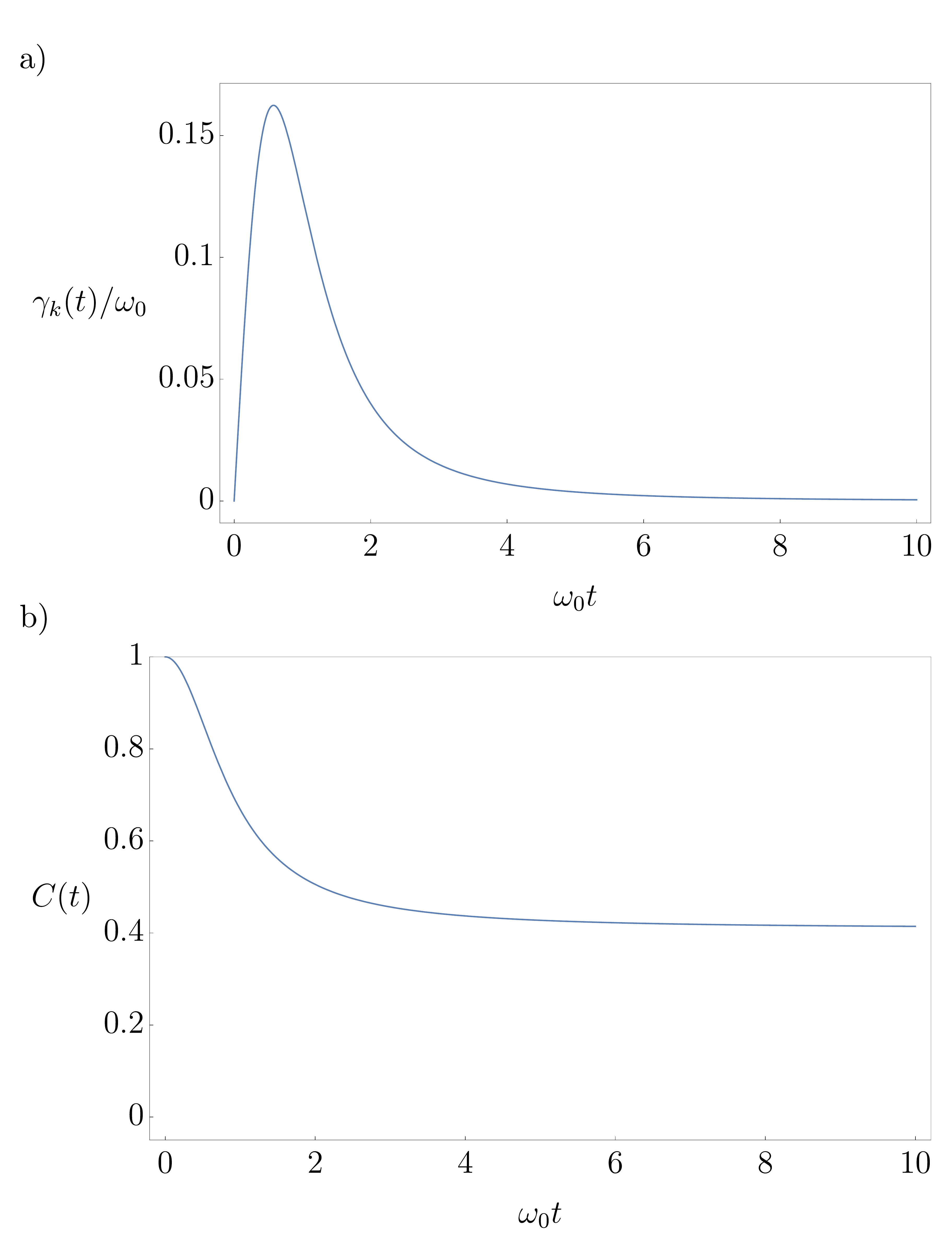}%
\caption{\label{NEWFIG2}(Color online) a) Time-dependent decay rate $\gamma_k (t)$ of Eq.~\eqref{GAMMAK} of the Pauli channel model,  with $s_k = 2$ and $\gamma_k = 0.25$. b) 
Time evolution of the associated CJ concurrence $C(t)$  of Eq.~\eqref{eq:9} with time-dependent decay rate. }
\end{center} 
\end{figure}

Following the same derivation as the previous section, the CJ state of the Pauli channel model can be written as 
{\small
\begin{eqnarray}
	&&\varrho^{AB}_{\Phi_{0\rightarrow t}}= 
	\\ \nonumber 
	&&\frac{1}{4}\begin{pmatrix}
	1+\beta_3(t) & 0 & 0 & \beta_1(t)+\beta_2(t)  \\
	0 & 1-\beta_3(t) & \beta_1(t)-\beta_2(t) & 0 \\
	0 & \beta_1(t)-\beta_2(t) & 1-\beta_3(t) & 0 \\
	\beta_1(t)+\beta_2(t) & 0 & 0 & 1+\beta_3(t)) 
	\end{pmatrix},
\end{eqnarray}}
with a CJ concurrence \eqref{NEWCON} equal to 
\begin{equation}\label{eq:9}
	C(t)=\max\{0,2p_0(t)-1 \},
\end{equation}
which we now study for 
some  paradigmatic examples of decaying rates $\gamma_k (t)$.
The first is obtained by considering the simple Markovian scenario where 
they are all taken to be non-negative constants, 
namely, $\gamma_k(t) = \gamma_k \geq 0$. In this case it is easy to see that Pauli channels always become EB after a certain characteristic length or time: 
indeed by direct inspection one notices that  $C(t)>0$ if and only if 
\begin{eqnarray} e^{-(\gamma_2 + \gamma_3)t} + e^{-(\gamma_3 + \gamma_1)t } + e^{-(\gamma_2 + \gamma_1)t} >1\;,\end{eqnarray} a condition
which is violated for sufficiently large $t$.
A completely different behavior is obtained 
instead by assuming the rates to be \begin{equation} \label{GAMMAK} 
	\gamma_k (t) = \gamma_k (1+t^2)^{-s_k / 2} \bar{\Gamma} (s_k ) \sin (s_k \arctan (t)),
\end{equation}
where $ \bar{\Gamma}$  is the Euler gamma function, $\gamma_k$ are positive coupling constants, $t$ is expressed in dimensionless units, and $s_k$ are the so-called Ohmicity parameters taking positive real values. These types of decay rates arise  from a microscopic model of a bosonic environment with an Ohmic-class spectral density (see, e.g., Ref. \cite {haikka-2013}), and they have been widely studied in the literature, mostly in the pure-dephasing dynamical case. Specific examples of the Ohmicity parameters $s_k$ are the case of the Ohmic environment $s_k=1$, sub-Ohmic environment $s_k<1$, and super-Ohmic environment $s_k>1$. It turns out that as long as $s_k\leq 2$, the model is still divisible, thereby exhibiting a Markovian character~\cite {haikka-2010}.
 In what follows we focus on the threshold case 
where the three decay rates are all equal to $2$, i.e., $s_k=s=2$.
Under this assumption the decay rates vanish after a finite time $\bar{t}$, and remain exactly zero thereafter; see Fig. \ref{NEWFIG2} a). As a consequence this system experiences entanglement trapping, as illustrated in Fig.~\ref{NEWFIG2} b), hence the channel is never EB, contrary to the case of positive constant decay rates. 

\begin{figure}[t!]
\begin{center}
\includegraphics[width=\linewidth]{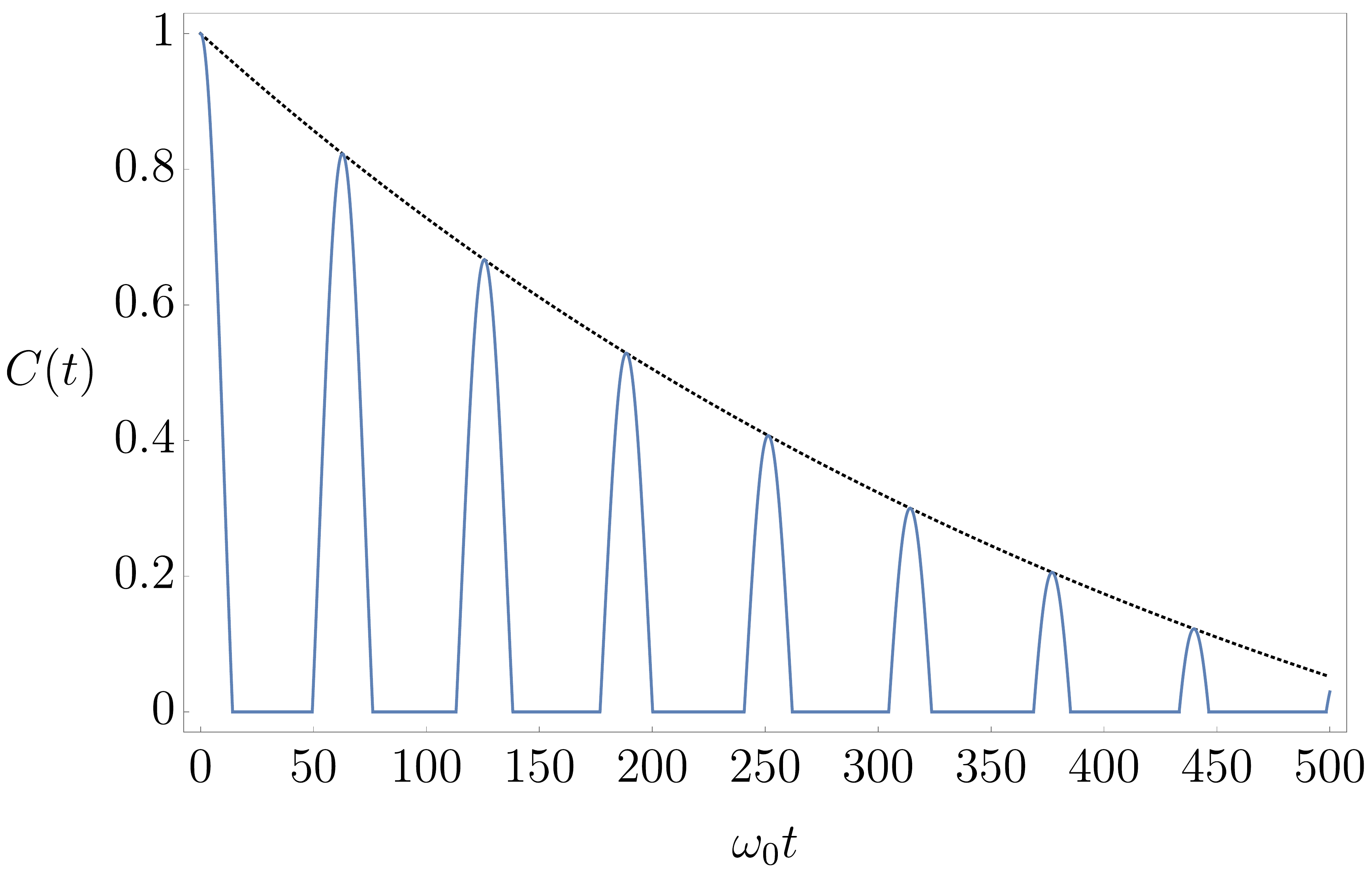}
\caption{\label{fig1}(Color online) CJ concurrence $C(t)$ for the Pauli channel model with uniform rates~\eqref{EPX} as a function of time in the non-Markovian regime  $2 \alpha_k / \lambda_k > 1$ (blue line) and the decaying envelope (black dotted line above).}  
\end{center}
\end{figure}

We now turn our attention to the case where the Pauli channel is non-Markovian, e.g., we consider as decay rates the functions 
\begin{eqnarray} \label{EPX} 
	\gamma_k(t)
= \frac{2 \alpha_k 
		}{\sqrt{1- \tfrac{2\alpha_k}{\lambda_k}}\; \mbox{cotanh} \left(\tfrac{\lambda_k t}{2} \sqrt{1-\tfrac{2 \alpha_k}{\lambda_k}}\right)+
	1} \;, 
	\end{eqnarray} 
with  $\lambda_k$ and $\alpha_k$ parameters associated to the microscopic details of the system-environment interaction. Note that, the master equation is nondivisible for $2 \alpha_k / \lambda_k > 1$, since in this case the decay rates take temporarily negative values. For simplicity we focus on the symmetric case $\gamma_1(t)= \gamma_2(t)= \gamma_3(t)$, such that $\alpha_k=\alpha$ and $\lambda_k=\lambda$ for $k=1,2,3$. Figure \ref{fig1} shows that the behavior of the concurrence in this regime is similar to that of the amplitude-damping channel (see Fig. \ref{fig0} for comparison). However, now there are extended intervals of transmission lengths for which entanglement is lost, while in the non-Markovian amplitude-damping case this happens only at certain times.

\section{Restoring entanglement via environment resetting} 
\label{SEC:4}

In this section we analyze what happens if during the system's evolution, as described by a family $\{ \Phi_{0\rightarrow t}\}_{t\geq 0}$, we allow for periodic  resetting of its environment.  
Specifically the idea is to divide the temporal axis 
 into a collection of 
time intervals ${\cal I}_n= [t_n, t_{n+1})$ which for simplicity we assume to have uniform length $\tau=t_{n+1}-t_n$ and $t_0=0$. 
Then at the end of each  interval we are assumed to  instantaneously reset the system 
environment  to  the  input state it had at the beginning of such an interval, essentially enforcing partial divisibility on the
system. 
The resulting  evolution of $A$ can  be described by a new $t$-parameter family
of perturbed mappings  $\{\tilde{\Phi}^{(\tau)}_{0\rightarrow t}\}_{t\geq 0}$  which for $t\in {\cal I}_{n}$ are defined by  the identity 
\begin{eqnarray} \label{nonloso}
	\tilde{\Phi}^{(\tau)}_{0\rightarrow t}&:=& 
	 \Phi_{0\rightarrow t-n\tau}\circ  \overbrace{( \Phi_{0\rightarrow \tau}) \circ \cdots \circ ( \Phi_{0\rightarrow \tau})}^\text{$n$  times} \nonumber \\ \nonumber \\ 
	 &=& 
	 \Phi_{0\rightarrow t-n\tau}\circ ( \Phi_{0\rightarrow \tau})^n\;. 
\end{eqnarray}
Our next step is to study how the EB properties of the problem get transformed in passing from $\Phi_{0\rightarrow t}$ to its perturbed counterpart 
$\tilde{\Phi}^{(\tau)}_{0\rightarrow t}$. We shall address this issue in the following subsections by applying the
construction  \eqref{nonloso}  to the qubit processes introduced in Sec.~\ref{SECiii}.
Before entering into this, however, we would like to make two remarks.

{\bf Remark 1:} From the divisibility analysis presented in 
Sec.~\ref{SECiiA}, it should be clear that  the map \eqref{nonloso} in general does not coincide with $\Phi_{0\rightarrow t}$. 
A notable exception is of course provided by time-homogeneous Markovian processes fulfilling the semigroup property~\eqref{DEFPRO0}
for which the equality 
\begin{eqnarray} 
{\tilde{\Phi}}^{(\tau)}_{0\rightarrow t} = {\Phi}_{0\rightarrow t} \label{QUESTA} 
\end{eqnarray} 
trivially holds for all choices of the interval length $\tau$. For these models no advantages or disadvantages can be expected from the periodic environment-resetting strategy. In this special scenario one could however modify  the
scheme \eqref{nonloso} by adding, for instance, periodic unitary rotations $U$ on $A$, along the line of the filtering
 scheme proposed in Refs.~\cite{anto,depasquale-2013,CUEVAS1,porzio}, creating the perturbed transformations 
\begin{eqnarray} \label{nonloso1}
	\tilde{\Phi}^{(U,\tau)}_{0\rightarrow t}	 &=& 
	 \Phi_{0\rightarrow t-n\tau}\circ ( {\cal U} \circ \Phi_{0\rightarrow \tau})^n\;,
\end{eqnarray}
where ${\cal U}(\cdot)=U\cdot U^\dag$ is a unitary channel; an example of this alternative approach is briefly presented in Appendix~\ref{APPEB}.
 However, the situation already changes
for  those Markovian processes which are not time-homogeneous: here,  due to the lack of the translational invariance property~\eqref{TRANS}, one has that the perturbed evolution  ${\tilde{\Phi}}^{(\tau)}_{0\rightarrow t}$ differs from the unperturbed one ${\Phi}_{0\rightarrow t}$.

{\bf Remark 2:}   Equation~\eqref{nonloso} admits a continuous limit when 
sending $\tau \rightarrow 0$ and $n\rightarrow \infty$ while keeping constant their product $n\tau\simeq t$, with $t - n \tau < \tau$. Indeed, assuming the maps $\Phi_{0\rightarrow t}$ of the original process 
 to be continuous and differentiable at the origin of the temporal axis,  we write  $\Phi_{0\rightarrow \tau} \simeq {\rm Id}+ \tau \; {\cal L}_0$, with ${\cal L}_0 =\tfrac{\partial}{\partial t}  \Phi_{0\rightarrow t}|_{t=0}$. Replacing this in Eq.~\eqref{nonloso} 
 we obtain
 $\tilde{\Phi}^{(t/n)}_{0\rightarrow t} \simeq
( {\rm Id} + \frac{t}n {\cal L}_0)^n$, and hence
 \begin{eqnarray} \label{CONT1} 
 \bar{\Phi}_{0\rightarrow t} := \lim_{n \rightarrow \infty}  \tilde{\Phi}^{(t/n)}_{0\rightarrow t} 
 = e^{{\cal L}_0 t} \;,
\end{eqnarray}
which is explicitly time-homogeneous and Markovian.

\subsection{Perturbed amplitude-damping channels} \label{PERAMP}

Let us first focus on the transformation \eqref{nonloso}  obtained when the unperturbed process $\Phi_{0\rightarrow t}$ is given by the amplitude-damping channel of Sec.~\ref{TLADP}. 
A simple  iteration of Eq.~\eqref{eq:ad-1q} reveals that in this case  $\tilde{\Phi}^{(\tau)}_{0\rightarrow t}$ is still
an amplitude-damping channel with a modified function $P(t)$. Specifically we have 
{\small
\begin{eqnarray}\label{eq:ad-1qMOD}
 \varrho^A(t) &=& \tilde{\Phi}^{(\tau)}_{0\rightarrow t}(\varrho^A(0))  \\
	&=& \begin{pmatrix}
	\varrho^A_{11}(0) \tilde{P}(t) & \varrho^A_{10}(0) \sqrt{\tilde{P}(t)} \\
	\varrho^A_{01}(0) \sqrt{\tilde{P}(t)} & \varrho^A_{00}(0) + \varrho^A_{11}(0)(1-\tilde{P}(t))
	\end{pmatrix},\nonumber 
\end{eqnarray}}%
where for $t \in {\cal I}_n$, the function $\tilde{P}(t)$ is obtained from $P(t)$ of the original process through the identity 
\begin{eqnarray}\label{FFF} 
\tilde{P}(t) = P(t-n\tau) P^n(\tau) \;,
\end{eqnarray} 
corresponding to a CJ concurrence   equal to 
\begin{equation}\label{CONCUPER}
	{C}(t) = \sqrt{\tilde{P}(t)}\;.\end{equation}
Notice that for $P(t)$ exponentially decreasing as in Eq.~\eqref{TIMEHO}, we have $\tilde{P}(t) = P(t)$, which implies  the identity \eqref{QUESTA} of Remark 1, 
in agreement with the fact that in this regime  the original process is time-homogeneous and
Markovian.
Regarding Remark 2 instead, we observe that in the present case, by direct computation,  the continuous limit process~\eqref{CONT1} is still an amplitude-damping channel of the form~\eqref{eq:ad-1q} with a probability parameter 
that is now given by
\begin{eqnarray} \label{NEWNEW}
\bar{P}(t) := e^{-\bar{\lambda} t} \;, \qquad \bar{\lambda}:=  - \frac{\partial}{\partial t} P(t)|_{t=0}\;.
\end{eqnarray} 

\begin{figure}[t!]
\begin{center}
\includegraphics[width=\linewidth]{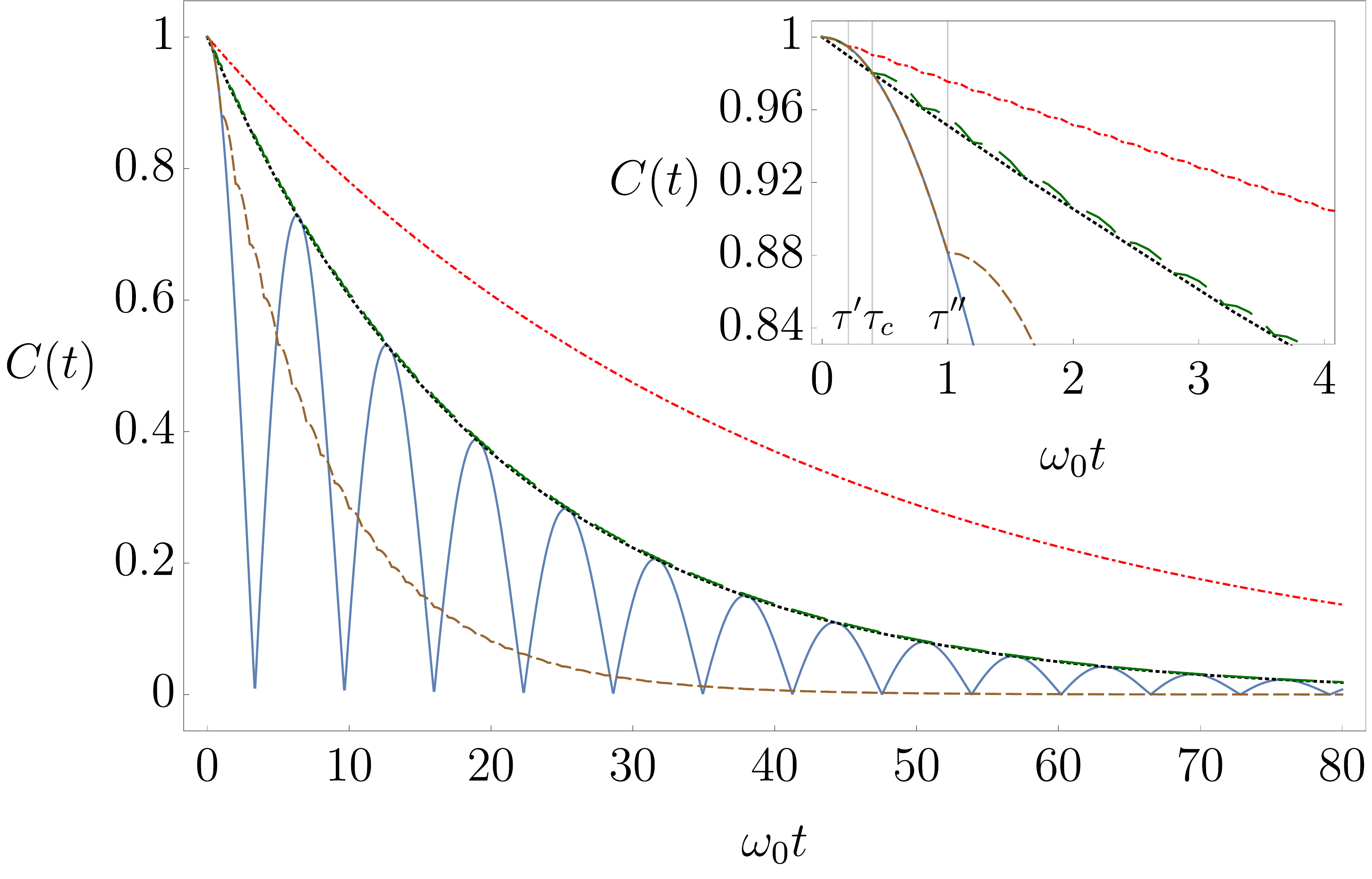}
\caption{\label{fig2}(Color online) Plot of the CJ concurrence \eqref{eq:sabb} of 
the amplitude-damping evolution for $\alpha = 5\omega_0$ and $\ell=0.1\omega_0$, 
as a function of time without interruptions to the dynamics (solid blue line) and the envelope of its peaks (dotted black line). The dynamics of CJ concurrence \eqref{CONCUPER}
for the perturbed map are depicted for $\tau'<\tau_c$ (red dot-dashed line), $\tau''> \tau_c$ (dashed brown line), and for $\tau=\tau_c$ (wider-dashed green line). The inset shows the initial behavior of these functions.}
\end{center}
\end{figure}

As an illustrative example we now assume $P(t)$ of the unperturbed model to be as in Eq.~\eqref{eq:poft}.
We have numerically observed that as long as $\tau$ is strictly smaller than  the value $\tau_1$ of Eq.~\eqref{ZEROS}, at which point the 
CJ concurrence of $\Phi_{0\rightarrow t}$   reaches the zero value for the first time, the perturbed CJ concurrence \eqref{CONCUPER} 
never vanishes, meaning that $\tilde{\Phi}^{(\tau)}_{0\rightarrow t}$ is prevented from reaching the EB regime at all times. 
By contrast, as soon as $\tau$ is at least as large as $\tau_1$ the perturbed channel acquires an EB character: in particular
for $\tau=\tau_1$, the family $\{ \tilde{\Phi}^{(\tau)}_{0\rightarrow t}\}_{t\geq 0}$ is EB for all $t\geq \tau_1$. 
Plots of the associated CJ concurrence \eqref{CONCUPER} of  $\tilde{\Phi}^{(\tau)}_{0\rightarrow t}$ 
are presented in Fig.~\ref{fig2} for various choices of the partition interval $\tau<\tau_1$ under the assumption the unperturbed
evolution is non-Markovian (i.e., $\alpha/\ell > 1/2$).
We have compared $\tau$ with the characteristic time $\tau_c = 1/\ell$ of the exponential decay for the unperturbed amplitude-damping model, which, for the
 sake of simplicity, we assume to be the smallest  time scale of the problem---a possibility that can be achieved by keeping  $\alpha/\ell-1/2$ positive but small. (Note that, for our considerations, $\tau_c<\tau_1$.)
 Under this circumstance one observes that 
if $\tau<\tau_c$ the CJ concurrence in the presence of interruptions (red line in the figure)  will always be greater than the one without interruptions, reaching the constant value of $1$ as $\tau$ approaches zero. This behavior 
can be understood by observing that for $P(t)$  as in Eq.~\eqref{eq:poft}, irrespective of the parameters $\ell$ and $\alpha$, we have 
\begin{eqnarray} 
\bar{\lambda}=  \frac{\partial}{\partial t} P(t)|_{t=0} =0, \Longrightarrow \bar{P}(t) =1\;,
\end{eqnarray} 
 implying that in the continuous limit \eqref{CONT1} the  perturbed transformation $\tilde{\Phi}^{(\tau)}_{0\rightarrow t}$
always approaches the identity channel:
 \begin{eqnarray} 
 \bar{\Phi}_{0\rightarrow t} = {\rm Id}\;.
 \end{eqnarray} 
 As we shall comment in the conclusions, this effect can be seen as a consequence of the Zeno effect, induced by the frequent resetting of its environment~\cite{facchi-2001,FacchiLidar}.
 
The preceding scheme could be implemented via recursive fast cooling of the two-level system; for example, the system and its local environment could be placed within a low (near-zero) temperature fridge that is directly connected to the latter. Within this setting, every $\tau$ seconds we could open the thermal contact for a brief period.

\subsection{Perturbed depolarizing channels}

\begin{figure}[t!]
\includegraphics[width=\linewidth]{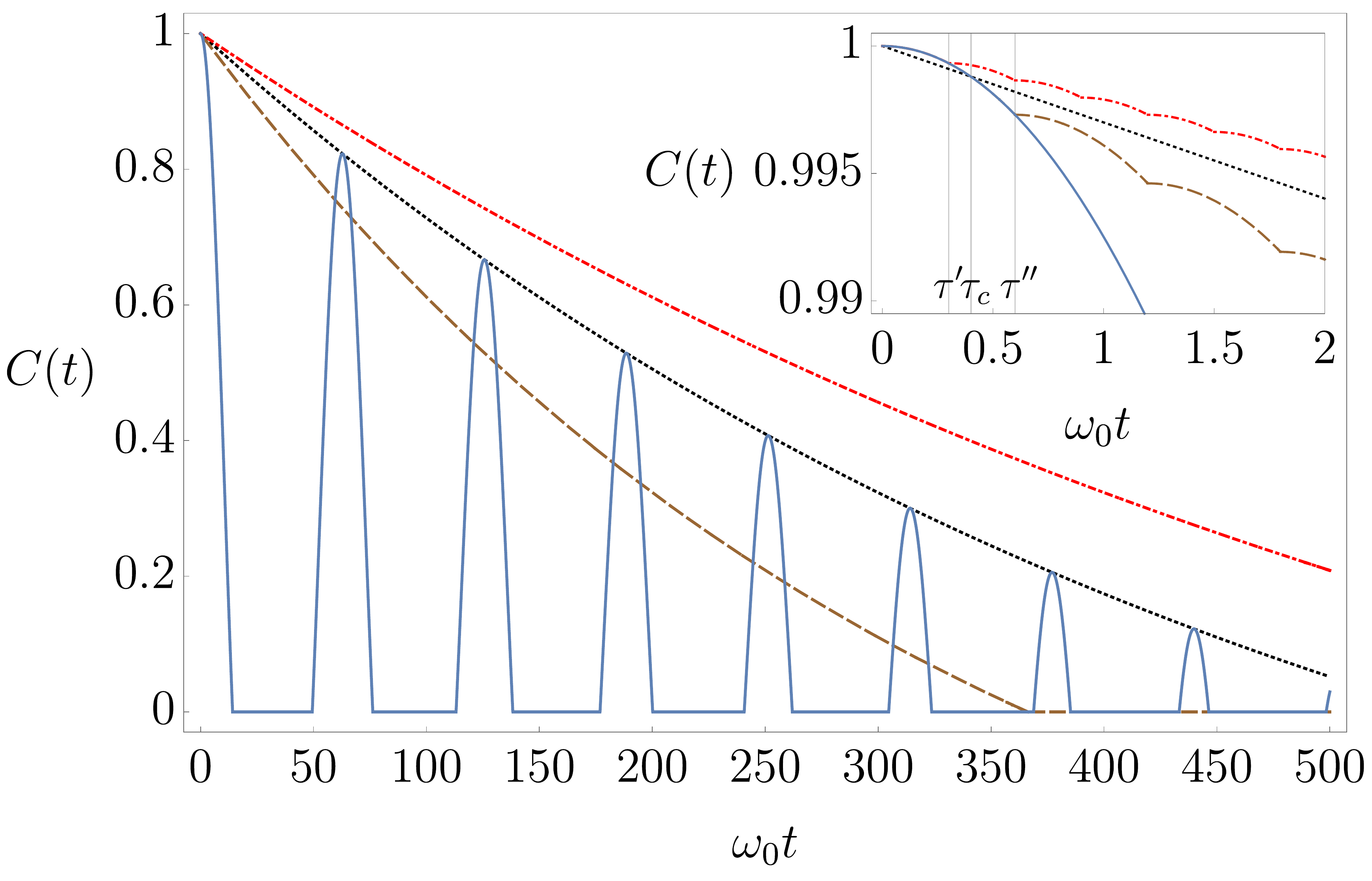}
\caption{\label{fig:2}(Color online) Temporal evolution of the unperturbed (blue solid curve) and perturbed CJ concurrences of a depolarizing channel, for two different values of $\tau$: $\tau' < \tau_c$ (red dot-dashed) and $\tau'' > \tau_c$ (brown dashed). The dotted black curve shows the decaying envelope of the unperturbed dynamical map. The inset contains a zoom-in of the short-time behavior. All curves have been produced assuming the 
rates of the Pauli channel as in Eq.~\eqref{EPX} with $\alpha_k=5\omega_0$ and $\lambda_k=10^{-3}$ for all $k$.}
\end{figure}

We now turn our attention to the case where the qubit system is subjected to a Pauli channel with a generator given by Eq. \eqref{eq:Paulit} focusing
on the symmetric case where the rates \eqref{EPX} are all identical leading to the depolarizing maps \eqref{eq:Pauli-uni}. 
By direct iteration one can easily verify that the perturbed map \eqref{nonloso} remains a depolarizing channel
with an effective noisy parameter, 
 \begin{eqnarray}\label{eq:Pauli-uniMOD}
	\tilde{\Phi}^{(\tau)}_{0\rightarrow t} (\varrho^{A}(0) ) 	&=& \tilde{\eta}(t)   \varrho^{A}(0)  + \frac{1-\tilde{\eta}(t)}{2}  \mbox{Tr}[   \varrho^{A}(0) ]\;, 
\end{eqnarray}
where for $t \in {\cal I}_n$ the function $\tilde{\eta}(t)$ is  obtained from the $\eta(t)$ of the original process \eqref{eq:Pauli-uni} through the same identity  we observed in Eq.~\eqref{FFF}:
\begin{eqnarray}\label{FFF1} 
\tilde{\eta}(t) = \eta(t-n\tau) \eta^n(\tau) \;.
\end{eqnarray} 
From this and from  Eqs.~\eqref{eq:9} and \eqref{NOISY}, the CJ concurrence of $\tilde{\Phi}^{(\tau)}_{0\rightarrow t}$ can
then be expressed in the following compact form:  
 \begin{equation} \label{FFD1}
C(t)=\mathrm{max}\Big\{0, \tfrac{1}{2}(3 \tilde{\eta}(t)-1) \Big\}.
\end{equation}
The continuous limit transformation \eqref{CONT1} can also be easily computed resulting once more in a
depolarizing channel with effective noisy parameter 
\begin{eqnarray}\label{FFF1NEW} 
\bar{\eta}(t) = e^{-\bar{\lambda}t}  \;, 
\end{eqnarray} 
with 
\begin{eqnarray} 
 \bar{\lambda}:= - \frac{\partial \eta(t)}{\partial t} \Big|_{t=0}=- \frac{4}{3} \frac{\partial  p_0(t)}{\partial t}\Big|_{t=0}=2\gamma(0)\;,
 \end{eqnarray} 
 where in the last two identities we used the identity~\eqref{NOISY}. 
Plots of \eqref{FFD1} are reported in Fig. \ref{fig:2} for the case where the original rate $\gamma(t)$  of the Pauli channel is as in Eq.~\eqref{EPX} revealing the analogous behaviors to those observed in the amplitude-damping case. Also, for this choice of $\gamma(t)$, one has $\gamma(0)=0$ and hence $\bar{\eta}(t)=1$ which implies 
again that the continuous limit of the perturbed map is given by the identity channel. 

\section{Discussion and Conclusions} \label{CONCLUSIONS}
In this paper we have studied the EB character of continuous-time quantum processes. The presence of non-Markovian effects causes  a nontrivial temporal dependence in the problem, as dynamics which  are EB for certain times  may become non-EB later on due to memory-induced entanglement revivals. In the spirit of reservoir engineering, we have also shown that EB properties can be manipulated by properly acting on the environment, e.g., via periodic resetting of its initial state. 
In particular, we have seen that in some cases, as the frequency of the perturbation increases, 
the dynamics get effectively frozen.  This is particularly interesting because it establishes a connection between our protocol and the quantum Zeno and inverse Zeno effects. It is nowadays agreed in the literature that the quantum Zeno effect (QZE) can be understood in a much more general framework than the one in which it was initially introduced. 
Indeed, this phenomenon was originally thought to arise due to the effect of frequent projective measurements on an open quantum system, where the measurements were performed  on the system of interest at intervals of time short enough to fall within the initial quadratic behavior characterizing its short-time dynamics. Later, it became clear that the QZE appears in a much broader context than its original formulation, namely, whenever a strong disturbance dominates the time evolution of the quantum system \cite{facchi-2001}. This is precisely what is mathematically described in Eq. \eqref{nonloso}, where the channel is considered as a concatenation of identical channels interrupted by unitary evolutions. This general description is, indeed, the one commonly used to study dynamical decoupling or bang-bang techniques, where instantaneous pulses (unitaries) are applied to an open quantum system to effectively decouple it from its environment. 
The connection between the quantum Zeno (and inverse Zeno) effects and dynamical decoupling has been thoroughly investigated in Ref.~\cite{FacchiLidar}, where it was shown that these dynamical phenomena can be seen as different manifestations of the same effect. In Ref. \cite {maniscalco-2008} it was shown that the quantum Zeno or inverse Zeno effect also affects the dynamics of entanglement, inhibiting or enhancing, respectively, its decay due to the interaction with the environment. In more details, depending on the properties of the system-environment interaction, there may exist a characteristic time $\tau_c$ such that, if measurements (or unitaries) are performed at time intervals $\tau < \tau_c$, then the entanglement decay is reduced, while if they are performed at intervals  $\tau >\tau_c$, then entanglement decay is enhanced, corresponding to Zeno and inverse Zeno effects, respectively. This is precisely what we have described in Sec. IV for both amplitude-damping and Pauli channels. While in Ref. \cite {maniscalco-2008}  both qubits were interacting with the environment, here we adopt the scenario typical of EB channels, i.e., we consider a qubit undergoing nonunitary evolution initially (maximally) entangled with an isolated ancilla. The effect of unitary interruptions, however, can be interpreted in the same spirit and it is similarly related to the quantum Zeno effect.

\begin{acknowledgments}
The authors from the Turku Centre for Quantum Physics acknowledge financial support from the Academy of Finland via the Centre of Excellence program (Project No. 312058) as well as Project No. 287750. T.B. acknowledges financial support from the Leverhulme Trust Scholarship {\it Study Abroad Studentship} (2016-2017) and from the Turku Collegium for Science and Medicine. A.D.P. acknowledges the financial support from the University of Florence in the framework of the University Strategic Project Program 2015 (Project No. BRS00215).
\end{acknowledgments}

\appendix
\section{Computing the CJ states}\label{CJSTATE} 

The CJ state~\eqref{CJEX1} for the amplitude-damping channel can be
easily derived, e.g., using the results of Ref.~\cite{bellomo-2007}, where 
 it was shown that if the reduced density matrices of a joint state  of the system $AB$ are expressed as
\begin{align}
	\varrho_{i i'}^A(t) &= \sum_{k k'} A^{k k'}_{i i'}(t) \varrho_{k k'}^A(0), \label{eq:Afunction}\\
	\varrho_{j j'}^B(t) &= \sum_{l l'} B^{l l'}_{j j'}(t) \varrho_{l l'}^B(0) \label{eq:Bfunction}, 
\end{align}
with some functions $A^{k k'}_{i i'}(t)$ and $B^{l l'}_{j j'}(t)$ with indices $i,i',j,j',k,k',l,l' = 0,1$, then the total system $AB$ has  matrix elements 
\begin{equation}
	\varrho_{ij,i'j'}^{AB}(t) = \sum_{k k' l l'} A^{k k'}_{i i'}(t) B^{l l'}_{j j'}(t) \varrho_{kl,k'l'}^{AB}(0), \label{eq:totden}
\end{equation}
(in the above expressions we use the symbol $\varrho_{k k'}^{X}$  to describe the matrix elements $\bra{k} \varrho^{X}\ket{k'}$). 
Applying this to evaluate the CJ state~\eqref{CJSTATE1}, Eq.~\eqref{CJEX1} follows by observing 
 that from Eq.~\eqref{eq:ad-1q} we have 
	\[
	\begin{split}
	A^{11}_{11}(t) &= 1-A^{11}_{00}(t) = P(t), \\
	A^{10}_{10}(t)&=A^{01}_{01}(t)=\sqrt{P(t)}, \\
	A^{00}_{00}(t)&=1,
	\end{split}
\]
and  $B^{ll'}_{jj'}(t) = \delta_{jl} \delta_{j'l'}$ as a consequence of the fact that no evolution is affecting the ancillary system $B$.

\section{Periodic environment resetting and filtering}\label{APPEB} 

\begin{figure*}[t!]
\subfloat[]{\label{fig2a}
  \includegraphics[width=.32\textwidth]{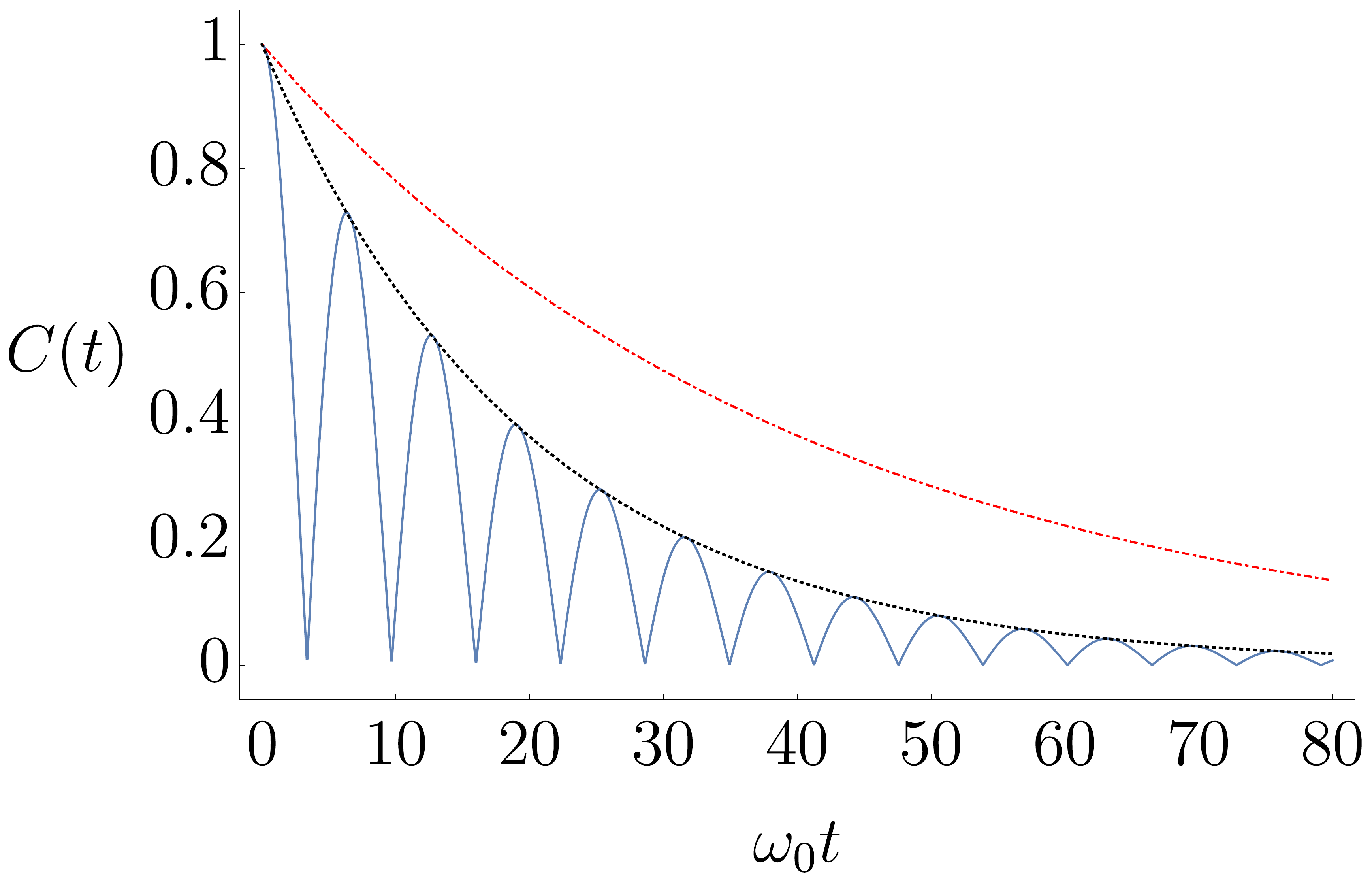}%
}\hfill
\subfloat[]{\label{fig2b}%
  \includegraphics[width=.32\textwidth]{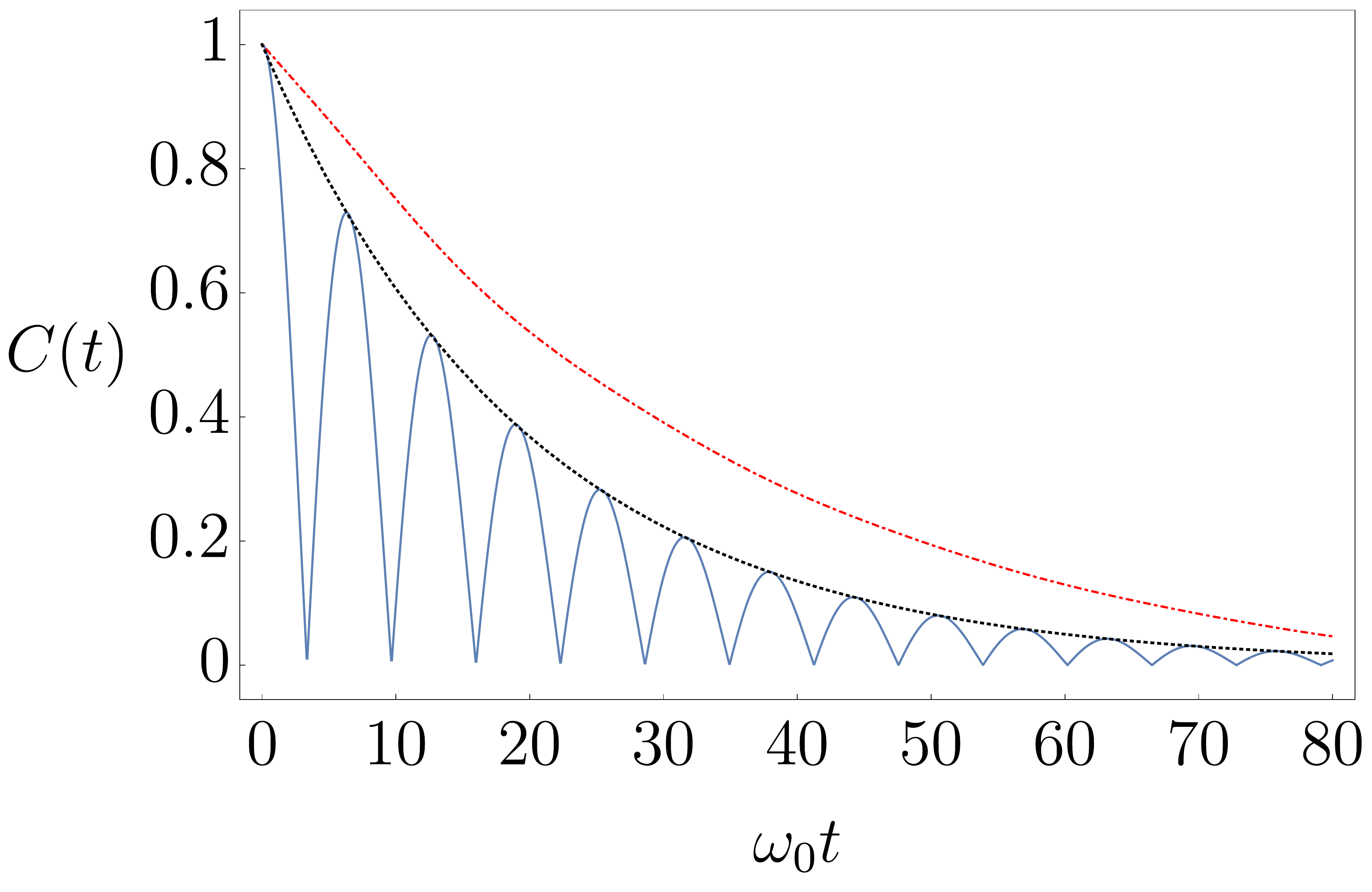}%
}\hfill
\subfloat[]{\label{fig2c}%
  \includegraphics[width=.32\textwidth]{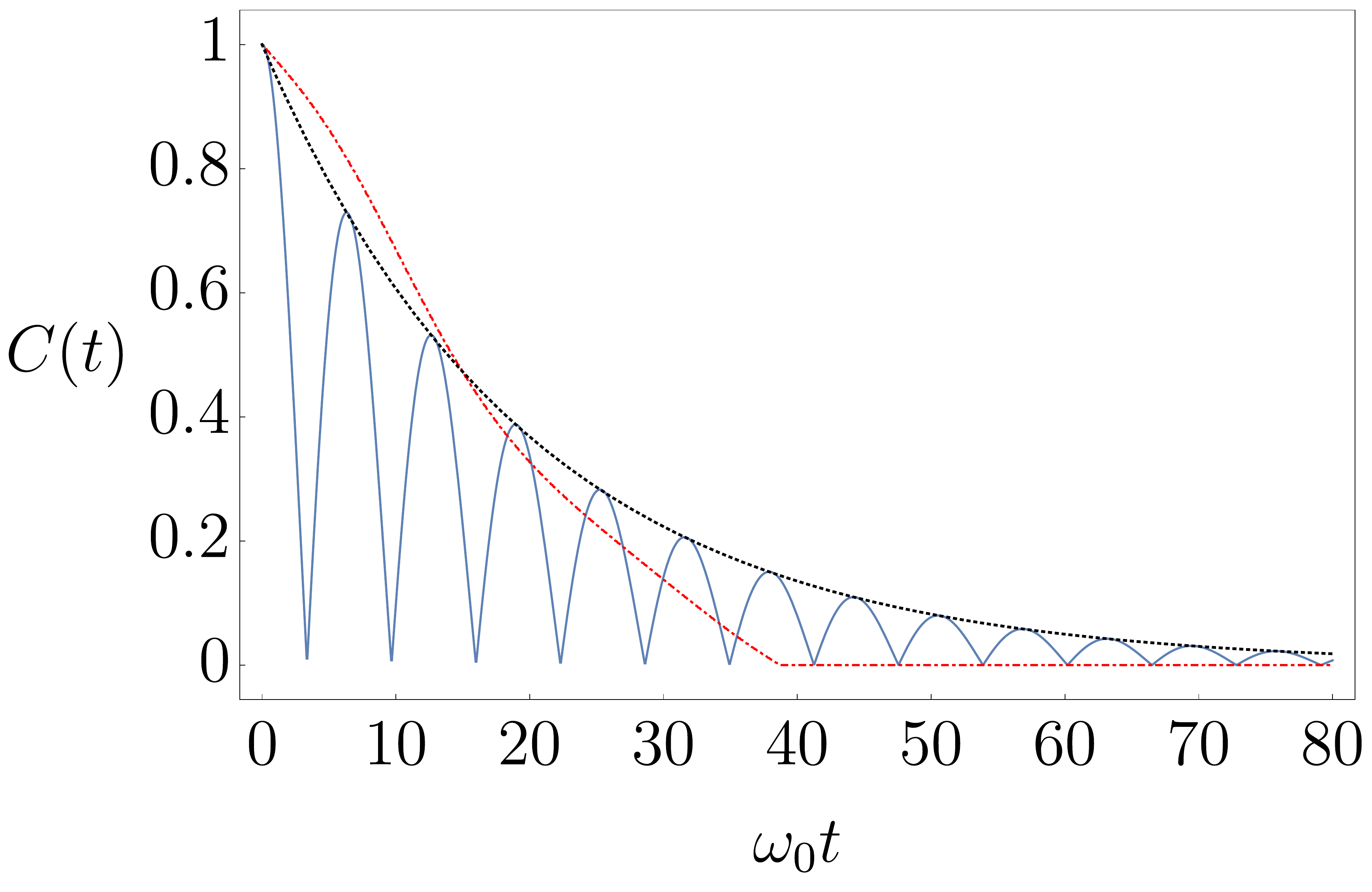}%
}
\caption{\label{fig6}(Color online) The effect of different instantaneous unitaries applied between the channels in the construction
of the perturbed channels of Eq.~\eqref{nonloso1}.
In each plot we show the CJ concurrence $C(t)$ of the unperturbed amplitude-damping processes ${\Phi}_{0\rightarrow t}$  without interruptions (solid blue line) and its
decaying envelope  (black dotted line), for $\alpha = 5\omega_0$ and $\lambda=0.1\omega_0$ as in Fig. \ref{fig2}. The temporal behavior of CJ concurrence of the associated perturbed maps \eqref{nonloso}
for different exemplary unitaries is shown by the red dot-dashed line: a) the unitary is the identity operator, i.e., $\theta=0$; b) $\theta=3.11$ and $\phi=0.5$; c) $\theta=3.11$ and $\phi=1.2$. The identity leads to optimal entanglement preservation (the same result 
has been verified numerically for other choices of the system parameters). Furthermore, as shown in 
panel c), some choices of unitary $U$ can lead to the complete and irreversible destruction of entanglement in the state.  }
\end{figure*}

Here we analyze the performances of the $U$-perturbed trajectories of Eq.~\eqref{nonloso1} 
for the amplitude-damping channel of Sec.~\ref{PERAMP}. Without loss of generality, we express
the unitary
as
	$U=   \openone \cos \theta+ i  \boldsymbol{r} \cdot \boldsymbol{\sigma} \sin\theta$,
[here $\boldsymbol{\sigma} = (\sigma_1, \sigma_2,\sigma_3)$ is the Pauli vector,  $\theta$ is a rotation angle, and 
	$\boldsymbol{r} = (\sin\phi, 0 , \cos\phi )$ fixes the rotation axis, the azimuthal angle being set equal to zero by exploiting the covariance of  the amplitude-damping process under rotation along the $z$ axis]. 
For a generic choice of the above parameters, a full analytical treatment of the problem
 produce results which are not particularly enlightening. For this reason we resorted  to a numerical analysis of the problem, reporting 
 our results in Fig.~\ref{fig6}. 


\bibliographystyle{unsrt}

\end{document}